\documentclass[useAMS,usenatbib]{mn2e}
\usepackage[cp866]{inputenc}
\usepackage[english]{babel}
\usepackage{graphicx}
\usepackage{amsmath}
\usepackage{amssymb}
\usepackage{epsf}
\usepackage{bm}

\def\la{\;\raise0.3ex\hbox{$<$\kern-0.75em\raise-1.1ex\hbox{$\sim$}}\;}
\def\ga{\;\raise0.3ex\hbox{$>$\kern-0.75em\raise-1.1ex\hbox{$\sim$}}\;}

\title[Quasi-periodical components]
{"Quasi-periodical components in the radial distributions of
cosmologically remote objects"}

\author[A.~I.~Ryabinkov, A.~D.~Kaminker]
{A.~I.~Ryabinkov$^{1}$, A.~D.~Kaminker$^{1, 2}$ \thanks{Send offprint request to: A.~D.~Kaminker} \\ 	 
$^{1}$ Ioffe Physical Technical Institute,
Politekhnicheskaya 26, 194021 St.~Petersburg, Russia,  \\
$^{2}$ St. Petersburg State Polytechnical University,  Politekhnicheskaya 29,\\
St. Petersburg, 195251 Russia, \\  
e-mail: calisto@rbcmail.ru, kam@astro.ioffe.ru}

\begin{document}

\date{Accepted 2011 xxxx. Received 2011 xxxx; 
in original form 2011 xxxx}

\pagerange{\pageref{firstpage}--\pageref{lastpage}} \pubyear{2010}

\maketitle

\label{firstpage}

\begin{abstract}
A statistical analysis of radial (line-of-sight)
1D-distributions of brightest cluster galaxies (BCGs)
within the redshift interval $0.044 \leq z \leq 0.78$
and Mg II absorption-line systems 
($0.37 \leq  z \leq 2.28$) is carried out. 
Power spectra and two-point radial correlation functions 
are calculated. It is found that both  radial
distributions of spectroscopic redshifts
of 52,683 BCGs and 32,840 Mg II absorption systems
incorporate similar quasi-periodical components 
relatively to the comoving distance.
Significance of the components exceeds 4$\sigma$-level and 
admits an increase ($\ga  5\sigma$) for
some broad  subsamples.   
For the $\Lambda$CDM cosmological
model the periodicities correspond
to spatial comoving scales
$(98 \pm 3)$  and $(101 \pm 2)$ h$^{-1}$ Mpc,
respectively. These quasi-periods turn out to be close
to the characteristic scale
$(101 \pm 6)$ h$^{-1}$ Mpc  
of the quasi-periodical component obtained earlier
for the radial distribution
of luminous red galaxies (LRGs). 
On the other hand,  the scales are close to the
spatial scale 
corresponding to the
baryon acoustic oscillations
(BAOs) revealed by many authors at the last decade.
Fourier transform phases obtained
for the BCGs and LRGs are found to be close,
while the phases calculated for the Mg II absorption systems
and LRGs are opposite. Discussions of the results in
a context of the BAO and large-scale structure 
characteristic scales are outlined. 
\end{abstract}


%
\section{Introduction}
\label{intro}
We continue a series of works on a research
of quasi-periodical scales in radial (line-of-sight) 
1D-distributions of various  cosmologically remote objects. 
The first two papers of this series 
(\citealt*{rkv07}; \citealt{rk11},
hereafter Papers~I and ~II) 
investigated quasi-periodical features
in the spatial-temporal distribution of 2003 and 2322 
absorption-line systems (ALSs) 
detected in the spectra of 661 and 730 QSOs, respectively, 
within the cosmological 
redshift interval $z=0.0$--4.3.   
In Paper~II we revealed
the existence of quasi-periodicity at a scale
$\Delta {\rm D}_c = (108 \pm 6)~h^{-1}$~Mpc,
where D$_c$ is a line-of-sight  comoving
distance to an ALS with absorption redshift 
$z_{abs}$. It was used the 
$\Lambda$CDM cosmological model with dimensional 
density parameter $\Omega_{\rm m}=0.23$.
The third  paper (\citealt*{rkk13}, hereafter Paper~III) 
analyzed the radial distribution of
106,000 luminous red galaxies (LRGs)
from the Sloan Digital Sky Survey (SDSS) 
catalogue, data release~7~(DR7),  
and found two most significant peaks
in its power spectrum. The peaks correspond
to the spatial comoving scales $(135 \pm 12)~h^{-1}$~Mpc
and $(101 \pm 6)~h^{-1}$~Mpc at the $\Lambda$CDM
model with $\Omega_{\rm m}=0.25$. The latter peak is
the dominant with significance $\ga 4\sigma$.
 
The appropriate scale of quasi-periodicity is 
in mutual accordance
with the period found for the radial distribution
of ALSs,  especially if one takes into account 
that at $\Omega_{\rm m}=0.25$ 
the spatial ALS-scale  shifts to    
$\Delta {\rm D}_c = (105 \pm 6)~h^{-1}$~Mpc. 
Moreover, both scales are in agreement 
with a spatial scale 
$(102.2 \pm 2.8)~h^{-1}$~Mpc
\citep{bl11},
which is widely accepted to be 
the scale of sound horizon, $r_s$, at the 
recombination epoch displaying itself
in cosmological galactic surveys as
the spatial scale corresponding to the baryon acoustic 
oscillations (BAOs;\,  
e.g.,   \citealt{eh98};  \citealt*{eht98};
\citealt{bg03}; \citealt{eisenetal05};
\citealt*{esw07}; \citealt{bh09}; \citealt{kaz10}; 
\citealt{anderetal12}; 
\citealt{ros12}, and references therein).   
To confirm statistically faintly outlined
relations between the BAO-phenomenon 
(mainly appearing in Fourier space)
and the quasi-periodical components
of the radial distribution of matter
it seems to be reasonable to examine
different types of
cosmologically remote objects 
in similar techniques.
This is one of the causes motivating 
us to carry out the present statistical
treatment.
      
In this paper we deal with the radial distributions 
of two independent sets of cosmological data:
(a) the spectroscopic redshifts $z$ of 52,683 
so-called brightest cluster galaxies (BCGs)
or the most luminous galaxies among
constituents of clusters
\citep{wh13} 
within the  interval $0.044 \leq z \leq 0.78$, 
and (b) the redshifts of 32,840 intervening 
Mg~II ALSs within the interval
$0.37 \leq z_{abs} \leq 2.28$ \citep{zm13}.
The updated catalog (a) of the BCGs 
sampled from 
spectral data of the SDSS DR9 catalog
\citep{ahnetal12}  
is available on the website %
\footnote{http://zmtt.bao.ac.cn/galaxy\_clusters};   
the  data (b) are based on the SDSS DR7 catalog   
(e.g. \citealt{abaz09})
and represented on the website%
\footnote{http://www.pha/jhu.edu/~gz323/jhusdss}.   
To carry out the statistical considerations  
we calculate the power spectra of radial distributions  
and separately --
two-point radial correlation functions (RCFs).   

The basic value of the power spectra calculations 
is a radial function N(D$_c$) 
integrated over angles $\alpha$
(right ascension) and $\delta$ (declination), 
D$_c (z)$ is the line-of-sight comoving distance
between an observer and cosmological objects under study;
N(D$_c$)dD$_c$ is a number of objects inside an interval 
dD$_c$. The radial comoving distances are calculated 
in a standard way  (e.g., \citealt*{khs97};  \citealt{h99})
\begin{equation}
{\rm D}_c (z_i) = {c \over H_0}\, \int_0^{z_i} 
{1 \over \sqrt{{\rm \Omega}_m (1+z)^3 + 
{\rm \Omega}_\Lambda}} {\rm d}z, 
\label{Dc}
\end{equation}
where $i$  numerates  redshifts $z_i$
of cosmological objects in a sample, 
$H_0=100~h$~km~s$^{-1}$ is the present Hubble constant,
$c$ is the speed of light, 
$c/H_0\ =\ 2998~h^{-1}$~Mpc;
it is accepted that
the dimensionless density parameters are
${\rm \Omega}_m = 0.25$ and 
${\rm \Omega}_\Lambda = 1-{\rm \Omega}_m = 0.75$. 
We use the binning approach and calculate so-called
normalized radial distribution function:
\begin{equation}
{\rm NN}({\rm D}_c) = {{\rm N}({\rm D}_c) - 
{\rm N}_{\rm tr}({\rm D}_c)
\over \sqrt{{\rm N}_{\rm tr}({\rm D}_c)}},
\label{NN}
\end{equation} 
where N$_{\rm tr}$(D$_c$) is a smoothed function  filtering
out the largest scales (trend function).
The denominator $\sqrt{{\rm N}_{\rm tr}}$ in Eq.~(\ref{NN})
stands for the standard deviation of the Poisson statistics.  
Examples of the trend function are represented in 
Fig.~\ref{N(D)}, where
the trends are calculated 
by the least-squares method with
using a set of parabolas, 
as regression functions for N(D$_c$),
and independent bins $\Delta_c = 10~h^{-1}$~Mpc.

Note that instead of the radial distribution function
N(D$_c$) one can use a comoving number density 
$n({\rm D}_c)$\ =\ N(D$_c$)/dV$_c$, where 
dV$_c$=$4\pi {\rm D}_c^2 {\rm d D}_c$ is a 
comoving differential
volume in the flat Universe, 
that is an analogue of the more conventional
value $n(z)$ (e.g., \citealt{zehavetal05};  \citealt{kaz10}).   
This replacement would not change Eq.~(\ref{NN}) 
and results of following calculations because in that
case one should divide 
both numerator and denominator 
by the comoving volume 
(remind that 
$\sigma(n_c)=\sigma({\rm N}_c)/{\rm d V}_c$,
where $\sigma$ is the standard (mean squared) deviation).

The power spectrum is calculated for 
the normalized radial   
distribution NN(D$_c$) 
according to the formula 
(e.g.,  \citealt{jw69}, \citealt{sc82}) 
\begin{eqnarray}
\hspace{1.0cm}
         {\rm P} (k_{\rm m}) & = & {1 \over {\cal N}_{\rm b}}   
                     \left\{ \left[
          \sum_{j=1}^{{\cal N}_{\rm b}}  {\rm NN}_j  
	  \cos(k_{\rm m} {\rm D}_{c, j})
              \right]^2   \right.
\nonumber                           \\
	      & +  &   
	   \left.    \left[ \sum_{j=1}^{{\cal N}_{\rm b}}  
	       {\rm NN}_j
             \sin (k_{\rm m} {\rm D}_{c, j})
             \right]^2 \right\},
\label{Pk}
\end{eqnarray}
where ${\cal N}_{\rm b}$ is a number of bins, 
$j=1,2,...{\cal N}_{\rm b}$ is a numeration of bins, 
${\rm D}_{c, j}$ is a location of bin centers,
$k_{\rm m}=2\pi {\rm m}/{\rm D}_c^{\rm L}$ is a wave number,
corresponding to an integer harmonic number
m$=1,2,...$,  ${\rm D}_c^{\rm L}$ 
is the whole comoving interval. 
Note that 
Eq.~(\ref{Pk}) represents just a square of  
the  discrete  Fourier
transform, $F_{\rm m}=F(k_{\rm m})$, 
responsible for the m-th harmonic
(e.g., \citealt{g75}):  
$|{\rm F}_{\rm m}|^2 = 
{\rm Re}^2({\rm F}_{\rm m}) + {\rm Im}^2({\rm F}_{\rm m}))$. 

We determine also the two-point 
RCF,
$\xi(\delta {\rm D}_c)$, for the samples under consideration
in an unconventional way (e.g., Papers~I and II). 
The variable $\delta {\rm D}_c$ is a radial comoving distance
between components of any pair of the objects in the samples.
Note that members of different pairs with fixed $\delta {\rm D}_c$ 
may correspond to different mutual comoving distances.
Using the estimator by \citet{ls93} one can introduce
\begin{equation}
\xi(\delta {\rm D}_c) = {{\rm DD} \over {\rm RR}} - 
                        2 {{\rm DR} \over {\rm RR}} + 1,
\label{Xi}
\end{equation}
where DD = DD$(\delta {\rm D}_c)$  
is a number of observed pairs of cosmological objects
separated by the radial comoving distance $\delta {\rm D}_c$
belonging to a range $\delta {\rm D}_c \pm \tilde{\Delta}_c/2$, where
$\tilde{\Delta}_c$ is a chosen accumulative bin width; 
DR = DR($\delta {\rm D}_c$) is a number of
cross pairs between points of the real sample and 
a random sample simulated  
for the same intervals of D$_c$ 
and  the same smoothed function (trend) as the real one;
RR = RR($\delta {\rm D}_c$) 
is a number of pairs counted up only for    
the random sample. 

We concentrate on the consideration of periodical
components  incorporated in the radial distributions 
of the samples  
relatively to the D$_c$-variable.
We treat these components as {\it quasi}-periodicities 
because of limited
intervals of line-of-sight distances used for 
both types of the objects,
as well as a dependence of peak 
positions and amplitudes 
on sample variations. 
In Sections~2 and 3 we present
the results of statistical analysis of galaxy clusters, 
traced by the BCGs 
and absorption systems Mg~II, respectively. 
In Section~4  we confront 
Fourier transform phases of the main 
quasi-periodical components with the phase
of the dominant quasi-periodical component
of the radial distribution of LRGs
(Paper~III). 
Conclusions and discussion of the results in a context
of the large-scale structure (LSS) 
of the matter distribution in the Universe 
are given in Section~5. 
 
\section{Radial distributions of brightest cluster galaxies}
\label{BCG}  
\begin{figure*}
\begin{center}
\epsfysize=55mm
\epsffile{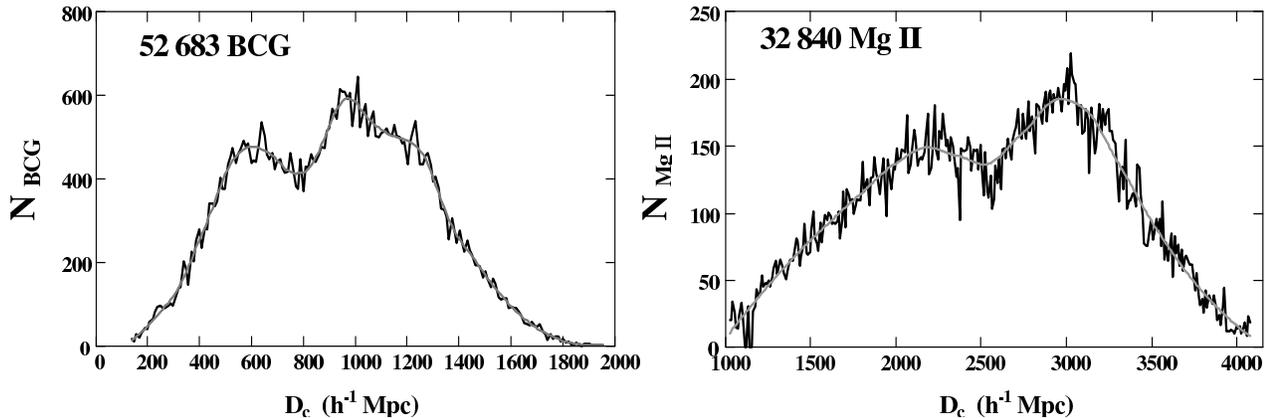}
\caption{
Radial distribution functions  
N(D$_c$), where D$_c$ is the line-of-sight 
comoving distance,  
calculated in $\Lambda$CDM cosmological model
with employing of independent bins $\Delta_c = 10$~h$^{-1}$~Mpc  
for two samples of cosmologically remote objects:
{\it left panel:}  52,683 brightest cluster galaxies (BCGs)
within the redshift interval $0.044 \leq  z  \leq 0.78$\  
($130.8 \leq {\rm D}_c \leq 1956.8$~h$^{-1}$~Mpc),
{\it right panel:} 32,840 absorption-line systems Mg~II
within the redshift interval $0.37 \leq z \leq 2.28$\ 
($1018.5 \leq {\rm D}_c \leq 4078.5$~h$^{-1}$~Mpc).   
Grey smoothed curves show the trend functions N$_{\rm tr}$(D$_c$). 
}  
\label{N(D)}
\end{center}
\end{figure*}

Fig.~\ref{N(D)} demonstrates 
the radial distribution functions  N(D$_c$) 
calculated for two samples of objects  
described in Introduction with using
independent bins  $\Delta_c = 10~h^{-1}$~Mpc. 
In both the cases  rather narrow spike-like 
variations of N(D$_c$) stand out against 
the background of the smoothed  trend 
functions N$_{\rm tr}$(D$_c$) (see Introduction) 
drawn by the grey curves. 

\begin{figure*}
\epsfysize=110mm
\epsffile{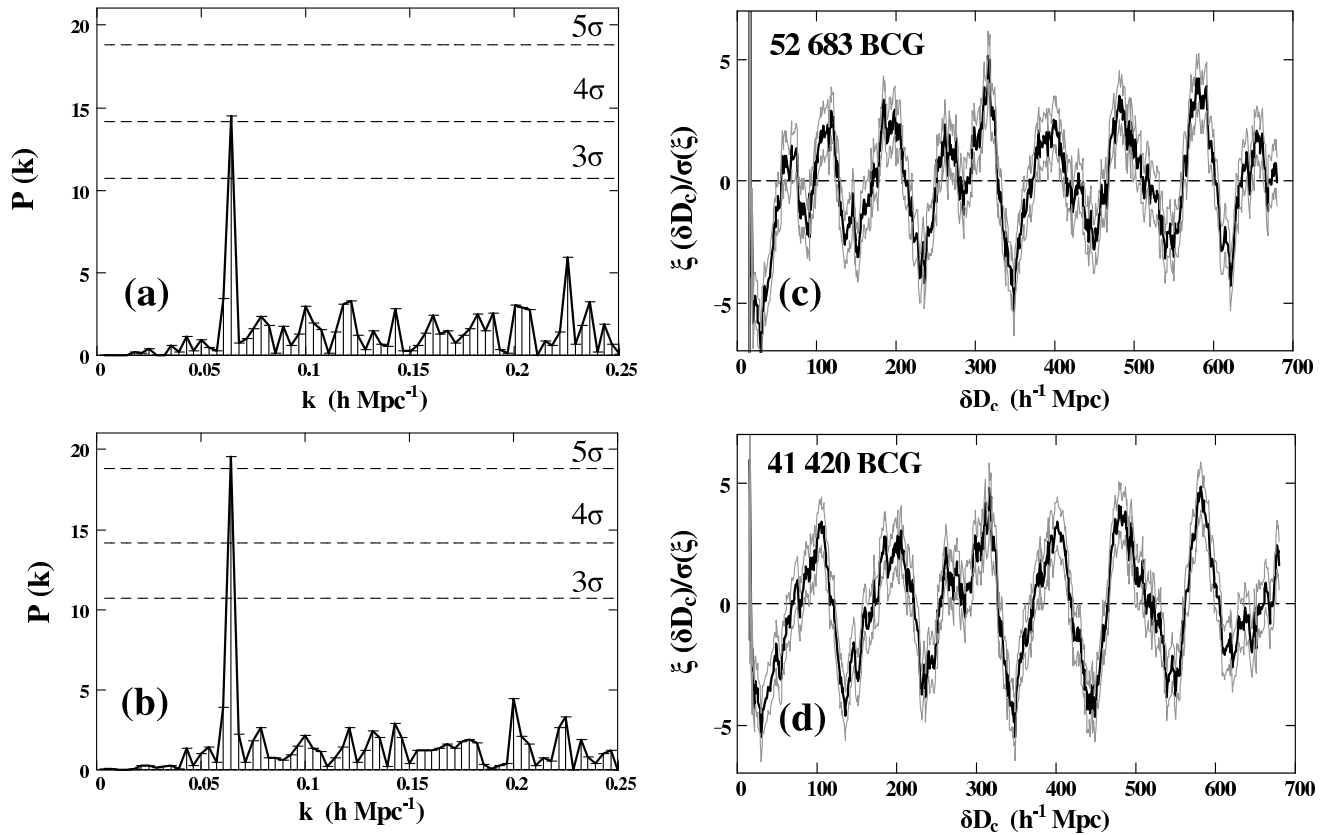}
\caption
{       
{\it Left panels} {\bf (a)} and {\bf (b)}:
power spectra P$(k)$ 
with $k = k_{\rm m}$ 
calculated 
according to Eq.~\protect{(\ref{Pk})}
for two BCG samples   
in the binning approach 
with independent bins $\Delta_c = 10$~h$^{-1}$~Mpc
within a  comoving-distance interval
$190.8 \leq {\rm D}_c \leq 1946.8$~h$^{-1}$~Mpc
(cf. Fig.~\ref{N(D)}) 
slightly reduced by a phase tuning procedure 
(see text); 
the horizontal dash lines specify the significance
levels $3\sigma$, $4\sigma$, and $5\sigma$ estimated 
with using  Eq.~\protect{(\ref{F})} at fixed levels
of confidence probability $\beta=1-{\cal F}=0.998$, 0.999936,
0.9999994, respectively. 
Panel {\bf (a)}:
P$(k)$ calculated for 
a slightly reduced number 52,526 of BCGs;           
panel {\bf (b)}: 
P$(k)$ calculated for the most luminous 
41,341~BCGs at the absolute magnitude 
$M_{\rm r}  \leq -23.01$.
{\it Right panels} {\bf (c)} and {\bf (d)}: 
normalized radial (two-point) correlation function (RCF)
(see text) calculated by sliding-average technique
with an accumulative bin  $\tilde{\Delta}_c = 30$~h$^{-1}$~Mpc 
and a step of its consecutive shifts $\delta_c=1$~h$^{-1}$~Mpc
in one of two directions from each point of the sample.
Grey curves indicate upper and lower boundaries 
of $\pm \sigma$ band of RCF-values. 
Panel {\bf (c)}: 
RCF calculated for the whole sample of 52,683~BCGs          
shown in the left panel 
of Fig.~\protect{\ref{N(D)}};
panel {\bf (d)}:  
RCF calculated for the most luminous 
41,420~BCGs at $M_{\rm r} \leq -23.01$  (see text).
}
\label{Pk_Xi_BCG}
\end{figure*}

Employing  Eqs.~(\ref{NN}) and (\ref{Pk})
one can calculate the  power spectrum 
for both the samples. 
Fig.~\ref{Pk_Xi_BCG}(a)
demonstrates
the resultant power spectra  P$(k)$
of the radial distribution of BCGs. 
A height of the single significant peak
has been slightly increased by 
a {\it phase tuning} procedure, i.e., 
a set of independent shifts of both the boundaries  
(D$_{c, {\rm min}}$ and D$_{c, {\rm max}}$) 
shortening the interval
to reach the visible
maximum of the amplitude P$_{\rm max}$
at m=m$^*$. 
In this way the most significant peak at 
m=m$^*$=18 or 
$k=k^*=0.0644~h$~Mpc$^{-1}$  is obtained 
for the interval
$190.8 \leq {\rm D}_c \leq 1946.8~h^{-1}$~Mpc
or D$_c^{\rm L}=1756~h^{-1}$~Mpc 
corresponding to the sample of 52,526~BCGs
(cf. 52,683~BCGs in Fig.~\ref{N(D)}).
The derived quasi-period is  
$\Delta {\rm D}_c = (98 \pm 3)~h^{-1}$~Mpc.

To estimate the significance of the main power spectrum peak 
we use hereafter so-called {\it false alarm probability} 
(see \citealt{sc82}; \citealt*{fef08} 
and references therein) treating all spectrum peaks
as a result of Gaussian noise:
\begin{equation}
{\cal F}={\rm Pr}({\rm P}_{\rm max} > {\cal P}_0) = 
1 - (1 - \exp(-{\cal P}_0))^{{\cal N}_q},
\label{F}
\end{equation}
i.e.,  the probability that  
an amplitude  P$_{\rm max}$ of
at least  one of possible 
peaks is higher than
a value ${\cal P}_0$.   
Here ${\cal N}_q$ 
is the Nyquist number,\  
${\cal N}_q = {\cal N}_{\rm b}/2$,
i.e., the maximal number of possible peaks
detectable in the power spectrum; 
in our case ${\cal N}_{\rm b}$ 
is determined in  Eq.~(\ref{Pk}) as the 
number of bins. 
Specifically  we have 
${\cal N}_{\rm b} = 175$~(BCGs)
and  ${\cal N}_{\rm b} = 293$~(Mg~II).
Choosing a given level of the false alarm probability
${\cal F}=p_0$ one can confront it with the amplitude ${\cal P}_0$
according to the formula 
${\cal P}_0=-\ln[1-(1-p_0)^{1/{\cal N}_q}]$,
or ${\cal P}_0=-\ln[1- \beta^{1/{\cal N}_q}]$,
where $\beta = 1-p_0$ is a confidence level. 
In this paper
we use the standard confidence levels
$\beta=0.998$ ($3\sigma$), 
$\beta=0.999936$ ($4\sigma$), and 
$\beta=0.9999994$ ($5\sigma$) 
for significance estimations.  
One can see that the significance of
the main peak in  Fig.~\ref{Pk_Xi_BCG}(a)
slightly exceeds the level $4\sigma$.
It evidences in favour of existence of a quasi-periodical
component in the radial distribution of BCGs.

\vspace{0.3cm}
Fig.~\ref{Pk_Xi_BCG}(b) represents the 
power spectrum calculations 
based on
a subsample 
of the 41,420~most luminous galaxies 
with  absolute magnitudes 
$M_{\rm r} \leq  M_0$ (index ``r'' stands for  SDSS ``r''-band)
selected from the whole sample of BCGs.
The upper limit $M_0$ was smoothly variated until
the main peak amplitude P$_{\rm max}$ in the power spectra
reached its maximal value at  $M_0 = -23.01$ 
with the significance exceeding the level
$5\sigma$. The power spectra in both 
left panels 
are calculated for the same  
interval of D$_c$ 
and in the same technique, 
but the panel (b) represents
a reduced number 41,341~BCGs. 
One can see strong enhancement of the peak
amplitude in the panel~(b) 
at the same m=m$^*$=18, $k=k^*=0.0644$, and 
$\Delta {\rm D}_c =(98 \pm 3)~h^{-1}$~Mpc
as in the panel~(a). 
That increases statistical reliability of 
the conclusion about occurrence of the quasi-periodical
component (signal) in the radial distribution of BCGs.
Note, however, that  
the estimation of the mean value  
of  relative oscillation amplitudes 
over the whole interval 
D$_c^{\rm L}=1826~h^{-1}$~Mpc
in this favourable case 
yields  a small number:\
$<\delta {\rm N}_{\rm BCG}/{\rm N}_{\rm tr, BCG}> \sim 0.04$. 

The right panels of
Fig.~\ref{Pk_Xi_BCG} display one-dimensional (two-point) 
normalized RCFs (see~Eq.(\ref{Xi})), 
$\xi(\delta {\rm D}_c)/\sigma(\xi)$,  
calculated by counting pairs 
(across the whole interval 
D$_c^{\rm L} = 1826~h^{-1}$~Mpc)
with fixed radial comoving  distance 
$\delta {\rm D}_c$. 
To estimate a standard  deviation 
$\sigma(\xi)$ for 
a set of the values
$\xi(\delta {\rm D}_c)$  
we  generate 
1000 random samples.   
It is found that 
the standard deviation 
can be approximated by  
$\sigma(\xi) \approx 1/\sqrt{\rm RR}$. 
In the panels~(c) and (d)  
the curves of grey colour 
indicate 
the bounds of the most probable 
quantities $\xi(\delta {\rm D}_c)$ 
restricted by $\pm \sigma$, i.e.,  
the quantities 
$\xi(\delta {\rm D}_c)/\sigma(\xi)$  
dispersed within a range $\pm 1$.    
The accumulative bin width is chosen as 
$\tilde{\Delta}_c=30~h^{-1}$~Mpc and 
a step of  consecutive shifts of
this bin is $\delta_c=1~h^{-1}$~Mpc.   

Fig.~\ref{Pk_Xi_BCG}(c) 
shows the normalized RCF 
of the whole sample 52,683 BCGs.
One can see a quasi-periodic behaviour of 
$\xi(\delta {\rm D}_c)$. If positions of the peaks 
(maxima) or depressions (minima) are determined 
as weighted mean values (centres of gravity, 
e.g., Paper~I) then the mean interval     
between neighbour peaks (or depressions)
matches with the period appeared  
in the power spectra.  
Some exception is the first peak located at 
$\delta {\rm D}_c \sim 50~h^{-1}$~Mpc. 
Such a peak in power spectra
appears for some realizations 
with minor significance ($\la 3\sigma$)
at  $k=0.120$ and 
$\Delta {\rm D}_c = 52.3~h^{-1}$~Mpc
in the course of 
the {\it phase tuning} procedure described above.
Let us note that similar quasi-period was
marked also in the radial distributions of the LRGs
(see Paper~III). 

The normalized RCF calculated for a sample
of the 41,420 most luminous ($M_{\rm r} \leq  -23.01$) 
BCGs is shown in Fig.~\ref{Pk_Xi_BCG}(d). 
In this case the oscillations
of the correlation function is 
more pronounced. The mean interval between 
centres of gravity of neighbour peaks (or depressions) 
is $(98.5 \pm 2.5)~h^{-1}$~Mpc, that is in a
good coordination with the results of spectral
analysis. Keeping this in mind          
we conclude that the most luminous BCGs
are the better tracers of the
quasi-periodic component in the radial
distribution of clusters than the whole 
sample of BCGs.
It seems to be concerned with 
more noisy statistics of the whole sample. 
 
\section{Radial distributions of Mg~II absorption-line systems}
\label{MgII}  
The right panel in Fig.~\ref{N(D)} demonstrates the radial 
distribution of 32,840  Mg~II ALSs within the interval
$1018.5~h^{-1} \leq {\rm D}_c \leq  4078.5~h^{-1}$~Mpc\
($0.37 \leq z_{abs} \leq 2.28$) plotted with the same 
independent bins $\Delta_c = 10~h^{-1}$~Mpc 
as it used  in the left panel. 
We consider only so-called intervening absorption
systems $z=z_{abs}$ 
blueshifted from their origin QSOs, $z_e$, 
at least by $\Delta z/(1+z_e) \geq \Delta v/c$,
where $\Delta z=z_e-z_{abs}$,  
$\Delta v=14,000$~km/s
is the minimal radial-velocity shift allowable 
for the absorption
systems relative to their $z_e$.
Catalog of the Mg~II absorption-line doublets 
cited in Introduction (see also \citealt{zm13}) 
contains the rest equivalent
widths  (at $\lambda = 2796 \AA$, first component)  
distributed in a wide range,  
$0.02 \leq {\rm W}_r(2796) \leq 8.4 \AA$.

\begin{figure*}
\epsfysize=110mm
\epsffile{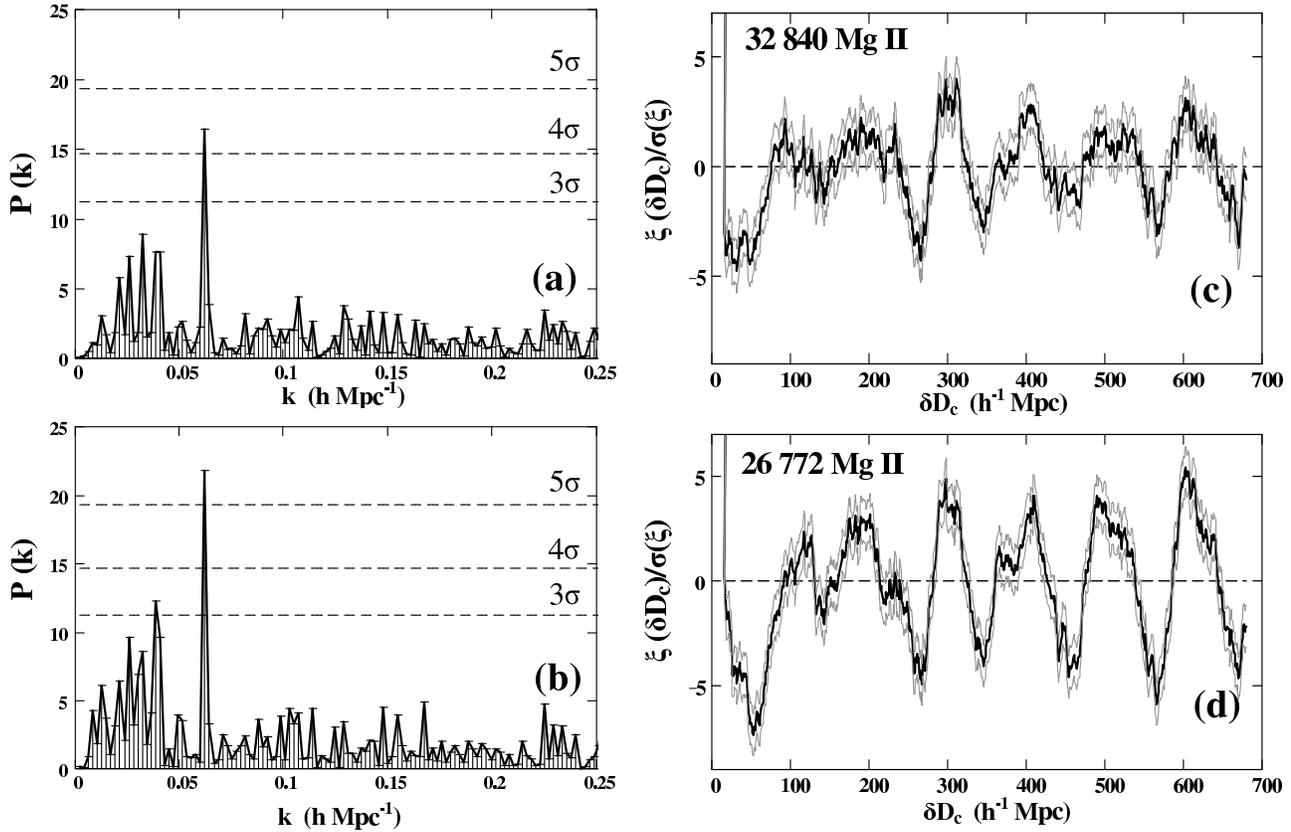}
\caption
{       
Same as in Fig.~\protect{\ref{Pk_Xi_BCG}}
but for two  samples of Mg~II ALSs.  
{\it Left panels} {\bf (a)} and {\bf (b)}: 
power spectra P$(k)$ calculated
in the same binning approach 
(independent bins -- $\Delta_c = 10$~h$^{-1}$~Mpc)
within a comoving-distance interval
$1128.5 \leq {\rm D}_c \leq 4058.5$~h$^{-1}$~Mpc
slightly
reduced by the phase tuning procedure (see text).
The confidence levels (horizontal dashed lines)
are also calculated  
with the use of Eq.~\protect{(\ref{F})}. 
Panel {\bf (a)}:  P$(k)$ calculated for 
a slightly reduced number 
32,589 of Mg~II systems;            
panel {\bf (b)}: P$(k)$ calculated for a subsample
of 26,561~Mg~II absorption systems selected with 
the use of a disjunction
of two conditions:  W$_r(2796) < 0.75\AA$\  $\bigvee$\ 
$\delta z \leq 0.0001$ at W$_r(2796) \geq 0.75 \AA$.    
{\it Right panels} {\bf (c)} and {\bf (d)}: 
same as in the panels {\bf (c)} and {\bf (d)} 
in Fig.~\protect{\ref{Pk_Xi_BCG}}
but for two  Mg~II ALSs samples obtained   
in the same sliding-average approach: 
the accumulative bin -- $\tilde{\Delta}_c = 30$~h$^{-1}$~Mpc 
and the step -- $\delta_c = 1$~h$^{-1}$~Mpc.
Panel {\bf (c)}:  RCF calculated for the whole sample of 
35,840~Mg~II  absorption systems involved in
the distribution represented in the right panel of 
Fig.~\protect{\ref{N(D)}};
panel {\bf (d)}:  RCF calculated for a subsample
of 26,772~Mg~II systems selected with the use of  
the same conditions as  
in the panel {\bf (b)}.  
}
\label{Pk_Xi_MgII}
\end{figure*}
The left panels in
Fig.~\ref{Pk_Xi_MgII} display 
the power spectra P$(k)$  plotted 
for the radial distributions of two 
Mg~II ALS samples  
in the same way as in Fig.~\ref{Pk_Xi_BCG}.
The panel (a) refers to a sample
of 32,589~Mg~II absorption doublets 
(cf. 32,840~Mg~II ALSs in Fig.~\ref{N(D)}) and 
reveals one dominant peak at
significance exceeding $4\sigma$.
The amplitude of the peak 
was also increased by 
the phase tuning procedure (described in Section~2). 
The confidence probabilities are calculated 
in the same way as in Fig.~\ref{Pk_Xi_BCG}(a) and (b) 
at a number of bins ${\cal N}_{\rm b}=293$. 
In this way the most significant peak at m=29,\ 
$k=0.0622~h$~Mpc$^{-1}$  is obtained 
for the interval
$1128.5 \leq {\rm D}_c \leq  4058.5~h^{-1}$~Mpc,
i.e.,  D$_c^{\rm L}=2930~h^{-1}$~Mpc. 
The  corresponding quasi-period is  
$\Delta {\rm D}_c = (101 \pm 2)~h^{-1}$~Mpc.

Fig.~\ref{Pk_Xi_MgII}(b) represents the 
power spectrum 
calculatins based on a subsample 
of 26,772~Mg~II systems submitting to 
two partly compatible   
conditions  W$_r(2796) < 0.75\AA$  or
$\delta z \leq 0.0001$ at W$_r(2796)  \geq 0.75 \AA$,
where $\delta z$ is an accuracy 
of the redshifts  $z_{abs}$ 
spectral measuring.   
These boundaries of the
values  W$_r$ and $\delta z$  have been fixed 
in a process of their smooth variations until
the main peak  P$_{\rm max}$
reached its maximal value  
with the confidence probability  exceeding
$5\sigma$. The power spectra in the 
panels~(a) and (b)
are also  (as in Fig.~\ref{Pk_Xi_BCG}) 
calculated  
for the same interval of D$_c$  and in the same 
technique, but the panel (b) represents  
a reduced number of 26,561 Mg~II systems.  
Let us emphasize that we get again  
the  strong enhancement of the peak
amplitude in the  panel~(b) 
at the same m=29, $k=0.0622$, and 
$\Delta {\rm D}_c =(101 \pm 2)~h^{-1}$~Mpc.  
In this case, as in Section~2,
the estimation 
of the mean value  
of  relative oscillation amplitudes 
over the whole interval 
D$_c^{\rm L}=3060~h^{-1}$~Mpc 
yields a small number
$<\delta {\rm N}_{\rm MgII}/{\rm N}_{\rm tr, MgII}> \sim 0.06$. 

The right panels in
Fig.~\ref{Pk_Xi_MgII} represent two-point RCF,   
$\xi(\delta {\rm D}_c)/\sigma(\xi)$,  
calculated by analogy with Fig.~\ref{Pk_Xi_BCG}
for the whole interval 
D$_c^{\rm L} = 3060~h^{-1}$~Mpc   
with the same accumulative bin width and 
the step of  consecutive shifts of the bin.      
In this case we also reproduce
1000 random samples and estimate the 
standard deviation for 
a resultant set of RCFs
using the same approximation 
$\sigma(\xi) \approx 1/\sqrt{\rm RR}$.
The grey coloured curves 
indicate again 
the bounds of the most probable 
quantities  
$\xi(\delta {\rm D}_c)/\sigma(\xi)$  
restricted by the standard deviations $\pm 1$.     

Fig.~\ref{Pk_Xi_MgII}(c)
shows a normalized RCF for the whole sample
of Mg~II doublets.  
The black and grey curves display 
a quasi-periodic behaviour of 
$\xi(\delta {\rm D}_c)$. 
The mean interval between 
centres of gravity of neighbour peaks 
(or  depressions)
turns out to be consistent with 
the period corresponding to the main
peak of the power spectra.

Fig.~\ref{Pk_Xi_MgII}(d)  demonstrates
the normalized RCF calculated for the sample
of  26,772~Mg~II ALSs 
satisfying the same two conditions 
as used in Fig.~\ref{Pk_Xi_MgII}(b). 
Similar to Fig.~\ref{Pk_Xi_BCG} one can see an increase
of the oscillation amplitudes.
The mean interval between 
centres of gravity of neighbour peaks  
proves to be  $(101 \pm 2)~h^{-1}$~Mpc, 
that is also in
good coordination with the results of the spectral
analysis. 
  
\section{Phases of quasi-periodical components}
\label{Phase}  
\begin{figure*}
\epsfysize=55mm
\epsffile{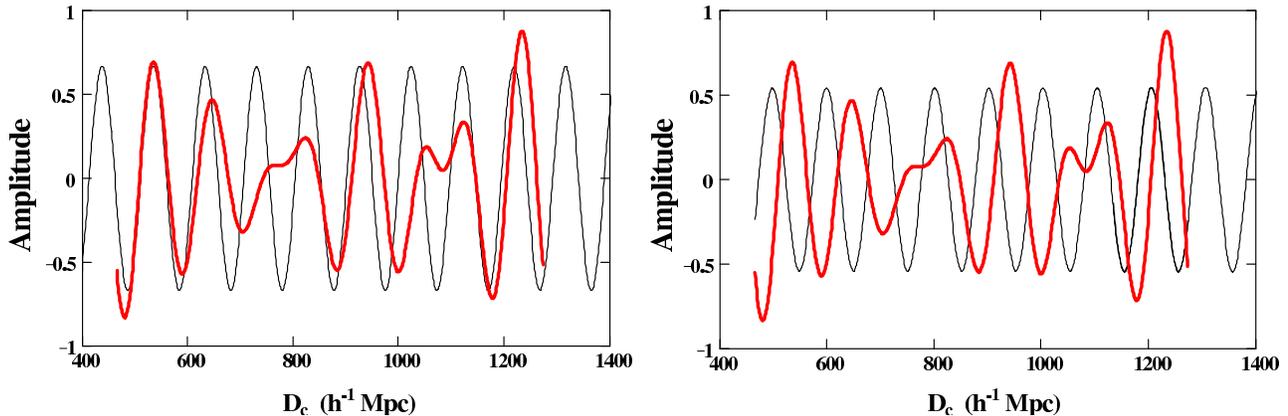}
\caption
{       
(Colour online) Reciprocal Fourier transformation 
produced for five harmonics  
$5 \leq {\rm m} \leq 9$ 
calculated in Paper~III
for the radial distribution 
of 105,831~LRGs (thick solid lines) versus two
reciprocal Fourier components (thin lines) 
calculated in the present work. 
{\it Left panel}: harmonic number 
m=18,\   ($\Delta {\rm D}_c = (98 \pm 3)$~h$^{-1}$~Mpc)
corresponding
to the single significant peaks
in  Figs.~\protect{\ref{Pk_Xi_BCG}}{\bf (a)} and {\bf (b)}.  
{\it Right panel}: harmonic number m=29\ 
($\Delta {\rm D}_c = (101 \pm 2)$~h$^{-1}$~Mpc)
corresponding to the single significant 
peak in Figs.~\protect{\ref{Pk_Xi_MgII}}~{\bf (a)} and {\bf (b)}.  
}
\label{Phases}
\end{figure*}

Results represented in Sections~2 and 3 are compatible 
with existence of the most significant  
quasi-periodical component 
at a characteristic scale  
$(101 \pm 6)~h^{-1}$~Mpc
in the radial distribution of LRGs  (Paper~III)    
mentioned in Introduction.  
Fig.~\ref{Phases} demonstrates 
a comparison
of phases of the Fourier transformations produced 
for the LRGs from one side and for BCGs 
and Mg~II ALSs  
from the other.  
In both panels of Fig.~\ref{Phases}
the same thick solid line  represents a sum of 
five harmonics calculated as 
the reciprocal Fourier transformation
of the radial distribution of LRGs
at $5 \leq {\rm m} \leq 9$.

Specifically, for the radial distribution of the LRGs 
we use an approximate expression:
\begin{equation}
{\rm NN}({\rm D}_c)  \approx {2 \over \sqrt{{\cal N}_{\rm b}}}\, 
                              \sum_{\rm m_{min}}^{\rm m_{max}}\,
                              |{\rm F}_{\rm m}(k_{\rm m})| \,
		               \cos  ( k_{\rm m}{\rm D}_c - 
		               \varphi_{\rm m}),
\label{NN-RF}
\end{equation} 
where  ${\rm F}_{\rm m}(k_{\rm m})$ is
the discrete Fourier amplitude 
introduced in the comments to Eq.~(\ref{Pk}),
$|{\rm F}_{\rm m}|= \sqrt{{\rm P} (k_{\rm m}})$,\, 
$k_{\rm m}=2\pi {\rm m}/{\rm D}_c^{\rm L}$ is a wave number and
$\varphi_{\rm m}= 
\arctan ({\rm Im}\ {\rm F}_{\rm m}/{\rm Re}\ {\rm F}_{\rm m})$
is a phase of m-th harmonic;  in our case
${\rm m_{min}} = 5$ and ${\rm m_{max}}=9$. 
Note that the main contribution
to the curve is provided by two harmonics:
m=6,\  $(135 \pm 12)~h^{-1}$~Mpc 
and the dominant at m=8,\  $(101 \pm 6)~h^{-1}$~Mpc.
                 
One can see that the mutual phase relations in the 
left and right panels of  Fig.\ref{Phases} 
are different. In the left panel
the thick curve is compared with a thin line or
the single spatial oscillation at m=18
corresponding to
the peak in the power spectrum
obtained for   
the most luminous 41,341~BCGs 
(Fig.~\ref{Pk_Xi_BCG}b).
In the right 
panel the thick curve is confronted with 
the single spatial oscillation at m=29 (thin line)
corresponding to the peak in the power spectrum
obtained for the  
26,561~Mg~II~ALSs 
(Fig.~\ref{Pk_Xi_MgII}b).
Both the spatial oscillations are calculated
with the use of  Eq.~(\ref{NN-RF})
at the fixed m.  

In the  left panel the phases 
of the main quasi-periodical
components of the LRGs and BCGs are close. 
Thus LRGs and 
galaxy clusters (or BCGs) trace the same 
elements of the large-scale structure (LSS) 
of cosmologically remote objects. 
It is not surprisingly because,
as it was emphasized by \citet{wh13},
majority of SDSS galaxy clusters includes  
LRGs as member galaxies
or the brightest galaxies inside clusters.    
Note, however, that the statistics of BCSs 
does not completely coincide with that
of LGRs.

In the right panel the phases 
of the main quasi-periodical components
calculated for 
the LRGs and Mg~II~systems 
turn out to be approximately opposite: 
mutually shifted at a phase angle $\sim \pi$. 
If these results were
confirmed it would mean that the 
LRG and Mg~II~ALS samples 
trace, as a rule,  
different elements of the LSS.   
For instance, if LRGs indicate regions with enhanced 
densities of dark matter then Mg~II~systems
concentrate in regions with reduced densities. 
This  assumption gets a 
qualitative support
in a visual pattern
of the combined 
LRGs and Mg~II~absorbers 
spatial distributions
represented
by  \citet{sretal08}. 
Although, the problem of spatial correlations 
of  Mg~II absorbers is considered to be  
rather complicated  
(e.g., \citealt{lundetal09}; \citealt{ms12}).

\section{Conclusions and  discussion}
\label{concl}
The main results of the 
statistical treatment  
of  52,683 brightest cluster galaxies (BCGs) 
from the SDSS DR9
within the redshift interval
$0.044  \leq  z  \leq  0.78$
\citep{wh13}
and 32,840
Mg~II absorption-line systems (ALSs)
from SDSS DR7 
within the interval 
$0.37 \leq z \leq 2.28$
\citep{zm13}
can be summarized as follows:  

(1)  The radial (line-of-sight)  
1D-distribution of  BCGs 
incorporates  a significant quasi-periodical 
component with respect to the smoother
radial selection function (trend).
The dominant peak in the power spectra
demonstrates the  periodicity 
corresponding to 
a spatial characteristic scale 
$\Delta {\rm D}_c = (98 \pm 3)~h^{-1}$~Mpc
at a level of significance $\ga 4\sigma$.
Still more prominent spectral
peak,  $\ga 5\sigma$, 
corresponding to the same scale
is obtained in the 
power spectrum calculated 
for a subsample 
of the 41,420~most luminous BCGs 
with absolute magnitudes $M_{\rm r} \leq -23.01$. 
Let us emphasize, however, that the mean value  
of  relative amplitudes of the oscillations 
averaged over the whole interval D$_c^{\rm L}$
for the latter subsample
turns out to be small  $\sim 0.04$. 

(2) The existence of the quasi-periodicity in 
the radial distribution of BCGs has been confirmed
also by the two-point  radial (1D) 
correlation function  
$\xi (\delta {\rm D}_c)$   
displaying the periodicity in the explicit form.

(3) The radial distribution of Mg~II~ALSs 
also includes a significant quasi-periodical component
with respect to the trend. The dominant peak in the
power spectra corresponds to a scale 
$\Delta {\rm D}_c = (101 \pm 2)~h^{-1}$~Mpc  
at the same significance level $\ga 4\sigma$ 
as for the BCGs. 
We also obtain    
an enhancement of the peak height
for the radial distribution 
of  26,772~Mg~II~systems separated out 
by two partly compatible   
conditions:  W$_r(2796) < 0.75\AA$  or
$\delta z \leq 0.0001$ at W$_r(2796)  \geq 0.75 \AA$,
where $\delta z$ is an accuracy of 
redshift $z_{abs}$ detections.
In the latter case 
the mean value  
of  relative amplitudes of the oscillations 
averaged over the whole interval  D$_c^{\rm L}$
is also small  $\sim 0.06$. 
Note that similar scale 
$\sim 100~h^{-1}$~Mpc
of the C~IV~ALSs 
correlation function 
within a range $1.2 < z  < 4.5$  
was found also by
\citet*{qvby96}.

(4) The two-point  radial (1D)  
correlation function  
$\xi (\delta {\rm D}_c)$ also visualizes
the quasi-periodical component 
in the radial distribution of Mg~II~ALSs
corresponding to the same scale as the
peak in the power spectra.  

(5) Both the quasi-periodical scales
found here in the radial distributions 
of two different types of cosmological
objects are consistent with the 
quasi-periodicity scale  
$(101 \pm 6)~h^{-1}$~Mpc 
found earlier (Paper~III) in the radial
distribution of the luminous red galaxies (LRGs). 
The phases of the main periodical components, 
calculated as 
reciprocal Fourier transforms, 
turn out to be approximately close 
for the BCGs and LRGs
and approximately opposite 
(shifted at a phase angle $\sim \pi$)
for the Mg~II~ALSs and LRGs. 

(6) All considered quasi-periodical 
scales (partly overlapping) 
are consistent with the scale 
$(102.2 \pm 2.8)~h^{-1}$~Mpc of the baryon
acoustic peak in the two-point spatial 
(3D-monopole) 
correlation function of the LRGs   
\citep{bl11}. 
It is likely that
there are implicit relations between
the quasi-periodicities found here in 
the radial distributions of cosmological objects
and the characteristic scale 
of sound horizon, $r_s$,  
at the recombination
epoch (so called  {\it standard ruler}; 
e.g., \citealt{bg03}; \citealt{percetal07}; \citealt{esw07}; 
\citealt{percetal10}; \citealt{anderetal12}).
In that case further 
confirmation and specification 
of the quasi-periodicity in radial distributions
would  become a simple and
alternative tool for an express-analysis 
of vast arrays of  cosmological data.  
On the other hand,   
the proximity of the scales of the baryon 
acoustic peak and those treated
in the present paper may evidence 
in favour of an assumption
that primordial acoustic
perturbations responsible for the BAO-scale 
could carry traces of a partly ordering 
formed at some early epochs, e.g., recombination
or radiation-matter equipartition.  

For instance, one can imagine
a weak  tendency of galaxies  
to be distributed as 
a continuous structure of higher and lower 
densities (a set of cells) formed by  
standing acoustic waves 
(e.g.,  \citealt{zn83},  \citealt*{hss97}, \citealt{esw07}) 
with dominant characteristic 
scale $\sim r_s$. 
This structure may 
display itself  in some apparent 
arrangement of the matter distribution,
as a consequence of a possible 
relatively dense packing
of the cells.  
In such a context 
the hypothesis
of a weak (linear) partial ordering 
of matter imprinted in the epoch 
of recombination 
(at $z \la z_{rec} \sim 10^3$)
may relate to
a conception of 
a quasi-regular network
formed by voids and galaxy superclusters  
at the nonlinear stage ($z \ll z_{rec}$),
which was discussed, e.g., 
in a series of papers by Einasto et al. 
(e.g.,  \citealt{einetal97a, einetal97b, einetal97c};
\citealt{ein00};
\citealt{tagoetal02}; 
\citealt{einetal11}).     

An illustration of the ordering possibilities 
was represented also
in Paper~II. There was simulated  
a model of partly ordered structure of points
(galaxies or galaxy clusters) forming 
a cloud-like scattering around the vertices 
of a simple cubic (SC) lattice. 
For this model it was produced
a mean power spectrum 
averaged over 100 random realizations
of the radial distribution function. 
The mean spectrum displays  
a significant 
peak  at $k=2\pi/l_{SC}$, where $l_{SC}$ 
is a lattice constant  and resembles
the  spectrum shown in Fig.~\ref{Pk_Xi_BCG}(a).
That reflects a general property of 
the radial distribution of the lattice points
with an arbitrary centre 
of the space. 
In this case the resultant radial distribution 
exhibits a periodical behaviour with a constant
amplitude of oscillations (e.g., \citealt{km63}).  

To elucidate the relation between the
oscillations observable in 3D-power spectrum, 
displaying the BAO phenomenon,  and 
quasi-periodical component in the radial distribution,
revealed in this work, we carried out some additional
simulations. 
To be specific, 
we produced a
square matrix of complex Fourier amplitudes 
$F(k_x, k_y)$ in a 2-dimensional $k$-space,
where $k_x$ and $k_y$ are     
two projections of the wave vector ${\bf k}$,
$k^2=k_x^2+k_y^2$, 
and Re$(F(k_x, k_y))$ and Im$(F(k_x, k_y))$
obey the Gaussian distributions, 
which are
modulated by harmonic variations
with smoothly decreasing (damped) amplitudes.  
The damping is similar to the observable 
baryon oscillations in 3D-$k$-space 
(e.g., \citealt{eh98};  \citealt{bg03};  
\citealt{anderetal12}).    
Then we transformed the generated 
in such a way   
square matrix of $F(k_x, k_y)$
into the distribution of points 
in the real 2D-space
and applied the analysis described 
in Introduction 
to random realizations of the distribution obtained.   

Our preliminary results have
shown that at some decreasing dependences of
the amplitudes at $k=k_1, k_2, k_3 ...$,
where $k_1, k_2=2 k_1, k_3=3 k_1 ...$ -- 
positions of the oscillation maxima,
the power spectrum  Eq.~(\ref{Pk}) 
of the radial distribution   
displayed a set of descending narrow peaks  
at the same $k=k_1, k_2, k_3 ...$.
Note that
subsequent  peaks cab be 
less significant than the first ones.
Moreover, the 
radial distribution
turns out to be 
sensitive to harmonic loops
of a 2D-$k$-distribution
transforming  them
to narrower peaks in the related 1D-$k$-space,
i.e., intensifying one (or a few) discrete 
radial harmonic(s).   
On the other hand, 
in the real 2D-space
we obtain only  
one noticeable peak of the standard (multi-central) 
correlation function. 
That reminds the well-known 
peak of the 3D-correlation function 
(e.g.,  \citealt{eisenetal05};  \citealt{kaz10}; 
\citealt{bl11};  \citealt{ros12})
specific for the BAO phenomenon in the real space.      
Being confirmed these results could 
reconcile  the baryon oscillations in the $k$-space
with quasi-periodicities of the radial
distribution in the real space.   

It is likely as well that 
the quasi-periodicities formulated in items (1)-(5)
concern  with the well known results by  
\citet{beks90} widened and confirmed by \citet{szetal93}.
Their pencil-beam surveys near both the Galactic poles 
within a redshift interval $z \la 0.5$
displayed a periodicity at a scale 128~$h^{-1}$~Mpc.  
Quasi-periodicities in 
two-dimensional patterns 
obtained on the base of 
the Las Campanas Redshift Survey data
had been also found by
\citet{landyetal96}  
as a  power-spectrum peak  
corresponding to a scale of
$\sim 100~h^{-1}$~Mpc 
in the spatial distribution of galaxies. 
This scale is rather 
close to the scales found here in the radial
distributions of the  BCGs and Mg~II ALSs,  
and found earlier for the LRGs.  
Possible relations of the cited periodicities
with the spatial scale of BAOs  
$\sim 100~h^{-1}$~Mpc were discussed by 
\citet{ehss98} and \citet{sz98}.\
\citet{szetal93} emphasized the special 
role of rich-statistics pencil-beam surveys 
in  bringing out the phase correlations (regularities) 
of matter density fluctuations. 
They expected  systematic 
decrease and broadening 
of the power spectrum peaks for wider 
angle regions of observational samples. 

Note, however, that      
the region of the SDSS DR9 survey
on the sky (e.g., \citealt{ahnetal12})
includes only the north galactic pole. 
Additional elimination from 
the sample of all objects 
within solid angles at 
$\theta \leq 5^\circ$ around the pole axis 
does not effect noticeably 
the radial quasi-periodical
component of the rest sample of BCGs.
Thus the effect of total radial distribution
considered here in a sense differs
from the effects discussed by \citet{beks90}
and \citet{szetal93}. Although, it is likely 
that there is 
some implicit relation between these two
approaches which still have to be manifested
in a future.

In reality quasi-periodical components of
the matter distribution may appear as a consequence
of two (or a few) factors, e.g.,  effects of clumping 
of galaxies and their clusters 
(e.g.,  \citealt{kp91})
and/or quasi-ordering
of the matter spatial distribution (discussed above). 
It is likely that the modern surveys of the sky 
deal with a complex mixture of both (or more) 
the factors.  In any case, 
the quasi-periodicity revealed here
being confirmed   
probably traces some spatial partly-ordered
structure imprinted in the LSS
of matter in the early Universe.  

\textit{Acknowledgments}
The work has been supported partly
by the
Research Program OFN-17, Division of Phys., RAS and 
by the State Program ``Leading Scientific 
Schools of Russian Federation'' (grant NSh-294.2014.2).




\end{document}